\documentclass[submission,copyright,creativecommons]{eptcs}
\providecommand{\event}{SR 2014} 
\usepackage{latexsym,amssymb,theorem,fleqn}

\newcommand{\tr}{\mathsf{t}}

\newcommand{\ATL}{\mathrm{ATL}}

\newcommand{\ATLA}{\mathrm{ATL}_\sqsubseteq}

\newcommand{\Subf}{\mathrm{Subf}}

\newcommand{\sem}[1]{\|#1\|}
\newcommand{\atlD}[1]{\langle\!\langle#1\rangle\!\rangle}
\newcommand{\atlB}[1]{[\![#1]\!]}

\newcommand{\until}[2]{{({#1}{\mathsf U}{#2})}}

\newcommand{\Act}{{\mathit{Act}}}

\newcommand{\out}{{\mathrm{out}}}

\newcommand{\asplit}[2]{\langle{#1}\sqsubseteq{#2}\rangle}
\newcommand{\bsplit}[2]{[{#1}\sqsubseteq{#2}]}
\newcommand{\Ag}{{\Sigma}}
\newcommand{\ran}{\mathrm{range}\,}

\newcommand{\oomit}[1]{}

\newtheorem{all}{Proposition}

{\theorembodyfont{\upshape}\newtheorem{definition}[all]{Definition}}
{\theorembodyfont{\upshape}}

\newenvironment{proof*}{\noindent
{\bf Proof:}}{$\dashv$}

\title{Refining and Delegating Strategic Ability in $\ATL$}
\author{Dimitar P. Guelev
\institute{Institute of Mathematics and Informatics\\
Bulgarian Academy of Sciences, Sofia, Bulgaria}
\email{gelevdp@math.bas.bg}
}
\def\titlerunning{Refining and Delegating Strategic Ability in $\ATL$}
\def\authorrunning{D. P. Guelev}
\begin{document}
\maketitle

\begin{abstract}
We propose extending Alternating-time Temporal Logic ($\ATL$) by an operator $\asplit{i}{\Gamma}\varphi$ to express that $i$ can distribute its powers to a set of sub-agents $\Gamma$ in a way which satisfies $\ATL$ condition $\varphi$ on the strategic ability of the coalitions they may form, possibly together with others agents. We prove the decidability of model-checking of formulas whose $\asplit{.}{.}$-subformulas have the form $\asplit{i_1}{\Gamma_1}\ldots\asplit{i_m}{\Gamma_m}\varphi$, with no further occurrences of $\asplit{.}{.}$ in $\varphi$.
\end{abstract}

\section*{Introduction}

The basic co-operation modality of Alternating-time Temporal Logics ($\ATL$, \cite{AHK97,AHK02}) invites perceiving agent coalitions as single agents who enjoy the combined powers of the coalition members. We investigate an operator to reverse this, by addressing the possibility to partition the strategic ability of a single agent among several sub-agents. We write $\asplit{i}{\Gamma}\varphi$ to denote that agent $i$ can partition its strategic ability among the members of a set of fresh sub-agents $\Gamma$ in a way which satisfies $\varphi$, a formula written in terms of the new agents $\Gamma$ who assume $i$'s powers, and the other original agents, except $i$. For example, a purchase scenario with the vendor represented by salesperson $SP$ and delivery team $DT$ can be described as
\[\asplit{\mathit{vendor}}{\mathit{SP},\mathit{DT}}
\left(\begin{array}{l}
\atlD{\mathit{customer},\mathit{SP}}\Diamond\mathit{purchase\ agreement}
\ \wedge\\
\atlB{\mathit{SP}}\Box(\mathit{purchase\ agreement}\Rightarrow\atlD{\mathit{DT},\mathit{customer}}\circ\mathit{delivery})\end{array}\right).\]
The combined powers of all of $i$'s sub-agents are always equal to $i$'s: \[\atlD{\Delta\cup\{i\}}\varphi\Leftrightarrow\bsplit{i}{\Gamma}\atlD{(\Delta\setminus\{i\})\cup\Gamma}\varphi\]
where $\bsplit{i}{\Gamma}$ stands for $\neg\asplit{i}{\Gamma}\neg$. Coalitions $\Delta\not\supseteq\Gamma$ may be weaker than $i$, but also have abilities contributed by agents from $\Delta\setminus\Gamma$. The realizability of schemes such as the example one generally depends on the basic composition of agents' actions. For instance, simple mechanisms make it always possible to deny the {\em proper} subsets of $\Gamma$ all substantial strategic ability or make $\Gamma$ use simple majority vote as indicated by the validity of the formula:
\[\neg\atlD{\emptyset}\varphi\wedge\atlD{i}\varphi\Rightarrow\asplit{i}{\Gamma}\bigwedge\limits_{\Delta\subsetneq\Gamma}\neg\atlD{\Delta}\varphi\wedge\asplit{i}{\Gamma}\bigwedge\limits_{\Delta\subset\Gamma,|\Delta|\leq|\Gamma\setminus\Delta|}\neg\atlD{\Delta}\varphi\wedge\bigwedge\limits_{\Delta\subseteq\Gamma,|\Delta|>|\Gamma\setminus\Delta|}\atlD{\Delta}\varphi.\]
Subtracting strategic ability from one agent and transfering it in the form of a virtual sub-agent to another is a way of implementing {\em delegation}. Refinement can be instrumental in expressing the {\em alienability} of the ability in question. E.g., 
\[\atlD{i}\circ\mathit{unlock}\wedge\neg\atlD{j}\circ\mathit{unlock}\wedge
\asplit{i}{i',\mathit{key}}
(\neg\atlD{i'}\circ\mathit{unlock}
\wedge\atlD{\underbrace{j,\mathit{key}}_{j'}}\circ\mathit{unlock})\]
states the possibility of giving $i$'s $\mathit{unlock}$ing ability separate identity $\mathit{key}$ which enables its passage to $j$. The relevant vocabulary introduced consists of $\mathit{key}$ itself, $\{j,\mathit{key}\}$ for $j$ $\mathit{key}$-in-hand and $i'$ for $i$ without $\mathit{key}$, respectively. 

Notably we investigate refining and delegating powers and not responsibilities as in, e.g., \cite{Norman:2002:GDR:544741.544856}. Sub-agents can pursue their own goals. As it becomes clear below, they do so by influencing the choice of actions on behalf of their super-agent with the share of the super-agents' power given to them. Unlike proper delegation as in, e.g., \cite{DBLP:journals/jair/HoekWW10} and \cite{DBLP:conf/sp/BandmannFD02}, where givers and receivers of control co-exist, just $\asplit{i}{\Gamma}$ is about {\em replacing} $i$ by its sub-agents $\Gamma$. 

Our main result about $\ATL$ with $\asplit{.}{.}$ in this paper is a model-checking procedure for the subset in which $\asplit{.}{.}$ is restricted to occur only in subformulas of the form $\asplit{i_1}{\Gamma_1}\ldots\asplit{i_m}{\Gamma_m}\varphi$, with no further occurrences of $\asplit{.}{.}$ in $\varphi$. This is sufficient for the handling of scenarios like the example one above, but with refinements affecting more than one primary agent.

\paragraph{Structure of the paper}  
After brief formal preliminaries on $\ATL$ on GCMs, we introduce our proposed operator and model-checking algorithm. We conclude by briefly commenting on some more related work, assessing our result and mentioning some work in progress. 

\section{Preliminaries}

\begin{definition}[concurrent game structures and models]
A {\em concurrent game structure} (CGS) for some given set of agents $\Ag=\{1,\ldots,N\}$ is a tuple of the form $\langle W,\langle \Act_i:i\in\Ag\rangle,o\rangle$ where 

$W$ is a non-empty set of {\em states};

$\Act_i$ is a non-empty set of {\em actions}, $i\in\Ag$; given a $\Gamma\subseteq\Ag$, $\Act_\Gamma$ stands for $\prod\limits_{i\in\Gamma}\Act_i$;

$o:W\times\Act_\Ag\rightarrow W$ is a {\em transition} function.

A {\em concurrent game model} (CGM) for $\Ag$ and atomic propositions $AP$ is a tuple of the form $\langle W,\langle \Act_i:i\in\Ag\rangle,o,V\rangle$ where $\langle W,\langle \Act_i:i\in\Ag\rangle,o\rangle$ is a CGS for $\Ag$ and $V\subseteq W\times AP$ is a {valuation relation}.
\end{definition}
In the sequel we always assume $\Act_i$, $i\in\Ag$ to be pairwise disjoint. 

Below we write $a_\Gamma$ to indicate that $a\in\Act_\Gamma$ where $\Gamma\subseteq\Ag$. If $a\in\Act_\Delta$ and $\Gamma\subseteq\Delta$, then $a_\Gamma$ also stands for the subvector of $a$ consisting of the actions for the members of $\Gamma$. Given disjoint $\Gamma,\Delta\subseteq\Ag$, we write $a_\Gamma\cdot b_\Delta$ for $c\in\Act_{\Gamma\cup\Delta}$ which is defined by putting $c_i=a_i$ for $i\in\Gamma$ and $c_i=b_i$ for $i\in\Delta$.

\begin{definition}[$\ATL$ on CGMs]
The syntax of $ATL$ formulas $\varphi$ is given by the BNF
\[\varphi,\psi::=\bot\mid p\mid (\varphi\Rightarrow\psi)\mid\atlD{\Gamma}\circ\varphi\mid\atlD{\Gamma}\until\varphi\psi\mid\atlB{\Gamma}\until\varphi\psi\]
where $p$ ranges over atomic propositions and $\Gamma$ ranges over finite sets of agents. Satisfaction of $\ATL$ formulas are defined in terms of strategies. A {\em strategy} for $i\in\Ag$ in CGM $M=\langle W,\langle \Act_i:i\in\Ag\rangle,o,V\rangle$ is a function from $W^+$ to $\Act_i$. Given a vector of strategies $s_\Gamma=\langle s_i:i\in\Gamma\rangle$ for the members of $\Gamma\subseteq\Ag$, the possible outcomes of $\Gamma$ starting from state $w$ and following $s_\Gamma$ is the set of infinite runs
\vspace{1.5mm}

\noindent
\ \ $\out(w,s_\Gamma)=\{w^0w^1\ldots\in W^\omega:w^0=w, w^{k+1}=o(w^k,a^k), a^0 a^1\ldots\in\Act_\Ag^\omega,\ a^k_\Gamma=s_\Gamma(w^0\ldots w^k), k<\omega\}$.
\vspace{1.5mm}

\noindent
Assuming a fixed $M$, we write $S_\Gamma$ for the set of all vectors of strategies for $\Gamma$ in $M$. Satisfaction is defined on CGMs $M$, states $w\in W$ and formulas $\varphi$:
\vspace{1.5mm}

\noindent
\ \ $\begin{array}{lll}
M,w\not\models\bot\\
M,w\models p &\mbox{iff}& V(w,p)\\
M,w\models\varphi\Rightarrow\psi &\mbox{iff}& \mbox{either }M,w\models\psi\mbox{ or }M,w\not\models\varphi\\
M,w\models\atlD{\Gamma}\circ\varphi & \mbox{iff} & \mbox{there exists an }s_\Gamma\in S_\Gamma\mbox{ s. t. }w^0w^1\ldots\in\out(w,s_\Gamma)\mbox{ implies }M,w^1\models\varphi\\
M,w\models\atlD{\Gamma}\until\varphi\psi & \mbox{iff} & \mbox{there exists an }s_\Gamma\in S_\Gamma\mbox{ s. t. for any }w^0w^1\ldots\in\out(w,s_\Gamma)\\
& & \mbox{there exists a }k<\omega\mbox{ s. t. }M,w^0\models\varphi,\ldots,M,w^{k-1}\models\varphi\mbox{ and }M,w^k\models\psi\\

M,w\models\atlB{\Gamma}\until\varphi\psi & \mbox{iff} & \mbox{for every }s_\Gamma\in S_\Gamma\mbox{ there exists a }w^0w^1\ldots\in\out(w,s_\Gamma)\\
& & \mbox{and a }k<\omega\mbox{ s. t. }M,w^0\models\varphi,\ldots,M,w^{k-1}\models\varphi\mbox{ and }M,w^k\models\psi
\end{array}$
\vspace{1.5mm}

\noindent
$\top$, $\neg$, $\vee$, $\wedge$ and $\Leftrightarrow$ and the remaining combinations of $\atlD{.}$ and $\atlB{.}$ with the temporal connectives $\circ$, $\Diamond$ and $\Box$ are regarded as derived constructs. See, e.g., \cite{AHK02} for the definitions.
\end{definition}

\section{Refining Strategic Ability in $\ATL$: $\ATLA$}

\begin{definition}[$\Gamma$-to-$i$ homomorphisms of CGMs]
Given $\Ag$ and $AP$, an $i\in\Ag$ and some non-empty set of agent names $\Gamma$ which is disjoint with $\Ag$, consider CGMs $M=\langle W,\langle \Act_j:j\in\Ag\rangle,o,V\rangle$ and $M'=\langle W',\langle \Act_j':j\in\Ag'\rangle,o',V'\rangle$ for $AP$, and $\Ag$ and $\Ag'=(\Ag\setminus\{i\})\cup\Gamma$, respectively. A mapping $h:\prod\limits_{j\in\Gamma}\Act_j'\rightarrow\Act_i$ is a {\em $\Gamma$-to-$i$ homomorphism from $M'$ to $M$}, if

$W'=W$, $V'=V$ and $\Act_j=\Act_j'$ for $j\in\Ag\setminus\{i\}$;

$\ran\,h=\Act_i$ and $o'(w,a)=o(w,a_{\Ag\setminus\{i\}}\cdot h(a_\Gamma))$ for all $w\in W$ and all $a\in\Act_{\Ag'}'$.

\end{definition}
Informally, if $M$ is a $\Gamma$-to-$i$ homomorphism of $M$, then the strategic ability of $i$ in $M$ is distributed among the new agents $j\in\Gamma$ in $M'$. For each action $a_i$ of $i$ in $M$ there exists a vector of actions $a_\Gamma$ for the members of $\Gamma$ in $M'$ such that $h(a_\Gamma)=a_i$. Together with the correspondence between the outcome functions $o$ and $o'$ of the two models, this means that the combined powers of the members of $\Gamma$ in $M'$ are equal to those of $i$ in $M$, but proper sub-coalitions of $\Gamma$ may be less powerful. Next we introduce the operator which is central to this work. Let $M$, $i$ and $\Gamma$ be as above. 
\begin{definition}[refinement operator]
Let $\varphi$ be written in terms of $(\Ag\setminus\{i\})\cup\Gamma$. Then
\[M,w\models\asplit{i}{\Gamma}\varphi\]
iff there exist an $M'$ for $\Ag'$ and $AP$ such that $M',w\models\varphi$, and a $\Gamma$-to-$i$ homomorphism from $M'$ to $M$.
\end{definition}
The occurrences of $j\in\Gamma$ in $\asplit{i}{\Gamma}\varphi$ are {\em bound} in the usual sense.
Informally, $\asplit{i}{\Gamma}\varphi$ means that $i$ can distribute its powers among the members of $\Gamma$ so that $\varphi$ holds in about the new set of agents. Its dual $\bsplit{i}{\Gamma}\varphi$ means that $\varphi$ holds regardless of how the powers of $i$ are distributed among the agents from $\Gamma$.

\section{Model-checking $\asplit{.}{.}^*$-Flat $\ATLA$}

$\asplit{.}{.}^*$-flat $\ATLA$ is the subset of $\ATLA$ in which  $\asplit{.}{.}$-subformulas have the form
\begin{equation}\label{asplitstarflat} \asplit{i_1}{\Gamma_1}\ldots\asplit{i_m}{\Gamma_m}\varphi
\end{equation}
where $\varphi$ has no further occurrences of $\asplit{.}{.}$. Note that only occurrences of $\asplit{.}{.}$ of the same polarity can be chained. E.g., if $\varphi$ and $\psi$ are $\asplit{.}{.}$-free, then $\atlD{i}\Diamond(\asplit{i}{\Gamma}\asplit{j}{\Delta}\varphi\wedge\bsplit{k}{\Upsilon}\bsplit{l}{\Xi}\psi)$ is $\asplit{.}{.}^*$-flat, but $\bsplit{i}{\Gamma}\asplit{j}{\Delta}\varphi$ and $\asplit{i}{\Gamma}\atlD{k}\Diamond\asplit{j}{\Delta}\varphi$ are not. Our algorithm reduces the model-checking problem to satisfiability in the $\atlD{.}\circ$-subset of $\ATL$, or, equivalently, in Coalition Logic \cite{Pauly02}, which is known to be decidable. We first do the case of $m=1$ and $\varphi$ being a boolean combination of $\atlD{.}\circ$-formulas with boolean combinations of atomic propositions as the arguments of $\atlD{.}\circ$, in full detail. Then we explain how the technique extends to arbitrary $m$, and, finally, however inefficiently, to formulas of the form (\ref{asplitstarflat}) with an $\asplit{.}{.}$-free $\varphi$ in which the use of the $\ATL$ connectives is unrestricted. 

\paragraph{The case of $m=1$} 
Consider some formula $\asplit{i}{\Gamma}\varphi$ with $\varphi$ restricted as above. Let CGM $M$ be as above and consider a CGM $M'=\langle W,\langle \Act_i':i\in\Ag'\rangle,o',V\rangle$, $\Ag'=\Ag\setminus\{i\}\cup\Gamma$, and a $\Gamma$-to-$i$ homomorphism $h$ from $M'$ to $M$. Let $\atlD{\Delta}\circ\chi$ be a subformula of $\varphi$. For $M',w\models\atlD{\Delta}\circ\chi$ to hold, there should be a vector of actions $a_\Delta$ such that, for any $b_{\Gamma\setminus\Delta}$, $a_{\Delta\setminus\Gamma}\cdot h(a_{\Delta\cap\Gamma}\cdot b_{\Gamma\setminus\Delta})$ gives $\Delta\setminus\Gamma\cup\{i\}$ a strategy to achieve $\circ\chi$ in $M$. For a fixed $a_{\Delta\setminus\Gamma}$ this means
\begin{equation}\label{actprops}
h(a_{\Delta\cap\Gamma}\cdot b_{\Gamma\setminus\Delta})\in
\{a_i\in\Act_i:\forall c_{\Ag\setminus(\Delta\cup\{i\})}M,o(w,a_{\Delta\setminus\Gamma}\cdot a_i\cdot c_{\Ag\setminus(\Delta\cup\{i\})})\models\chi
\}
\end{equation}
Henceforth we write $A_{i,a_{\Delta\setminus\Gamma},w,\chi}$ for the subset of $\Act_i$ in (\ref{actprops}).

Now consider a CGM $\overline{M}=\langle \overline{W},\langle \overline{\Act}_j:j\in\Gamma\rangle,\overline{o},\overline{V}\rangle$ for $\Gamma$ as the set of agents, $\overline{AP}=\Act_i$ as the set of atomic propositions and $\overline{W}=\Act_i\cup\{w^0\}$ as the set of states. Let $\overline{V}(w,a)$ be equivalent to $w=a$ for $a\in\Act_i$, thus enabling reference to each individual action of $i$. The intended meaning of the states of $\overline{M}$ from $\Act_i$ is to represent the possible choices of $i$'s actions by the members of $\Gamma$; $w^0$ is a distinguished reference state. Let $\overline{\Act}_j=\Act_j'$ for $j\in\Gamma$, and let $\overline{o}(w^0,a)=h(a)$ for all $a\in\overline{\Act}_\Gamma$. Then 
\begin{equation}\label{bool2}
\overline{M},w^0\models\atlD{\emptyset}\circ\bigvee\limits_{a\in\Act_i}a\wedge\bigwedge\limits_{a,b\in\Act_i,a\not=b}\atlD{\emptyset}\circ\neg(a\wedge b)\wedge\bigwedge\limits_{a\in\Act_i}\atlD{\Gamma}\circ a,
\end{equation}
since, due to the surjectivity of $h$, each of $i$'s actions can be enforced by $\Gamma$, which is the grand coalition in $\overline{M}$.

Let the translation $\tr$ replace subformulas of $\varphi$ of the form $\atlD{\Delta}\circ\chi$ by their corresponding \[\bigvee\limits_{a_{\Delta\setminus\Gamma}\in\Act_{\Delta\setminus\Gamma}}
\atlD{\Delta\cap\Gamma}\circ \bigvee\limits_{a_i\in A_{i,a_{\Delta\setminus\Gamma},w,\chi}}a_i.\] 
Then $M,w\models\asplit{i}{\Gamma}\varphi$ is equivalent to $\overline{M},w^0\models\tr(\varphi)$. 

Conversely, let a model $\overline{M}=\langle \overline{W},\langle \overline{\Act}_j:j\in\Gamma\rangle,\overline{o},\overline{V}\rangle$ exist such that $\overline{M},w^0\models\tr(\varphi)$ and (\ref{bool2}) hold. Then we can define an $M'$ and a $\Gamma$-to-$i$ homomorphism $h$ to witness $M,w\models\asplit{i}{\Gamma}\varphi$ as follows. We put $\Act_j'=\overline{\Act}_j$, $j\in\Gamma$. For every $a_\Gamma\in\overline{\Act}_\Gamma$, we define $h(a_\Gamma)$ as the unique $a_i\in\Act_i$ such that $\overline{M},o(w^0,a_\Gamma)\models a_i$. 
The identity $o'(w,a)=o(w^0,h(a))$ determines $o'$. Now a direct check shows that $M,w\models\asplit{i}{\Gamma}\varphi$.  

Hence, the existence of a model $\overline{M}$ which satisfies $\tr(\varphi)$ and (\ref{bool2}) at some state is equivalent to the satisfaction of $\varphi$ at the given state $w$ of the given $M$. Since satisfiability of formulas such as $\tr(\varphi)$ and (\ref{bool2}) is solvable, this entails the solvability of model-checking $\asplit{.}{.}$-formulas.

\paragraph{The case of $m>1$} To keep notation simple, let $m=2$, i.e., consider formulas of the form $\asplit{1}{\Gamma_1}\asplit{2}{\Gamma_2}\varphi$. Bigger $m$ are handled analogously. We first revise condition (\ref{actprops}), with respect to formulas $\atlD{\Delta}\circ\chi\in\Subf(\varphi)$ in which $\Delta\subseteq\Ag'$, $\Ag'=\Ag\setminus\{1,2\}\cup\Gamma_1\cup\Gamma_2$. The $m=2$-form of (\ref{actprops}) is about sets of {\em pairs} of actions, for $1$ and $2$, respectively. Given a fixed $a_{\Delta\setminus(\Gamma_1\cup\Gamma_2)}$, (\ref{actprops}) assumes the form
\[\begin{array}{l}
\langle 
h_1(a_{\Delta\cap\Gamma_1}\cdot b_{\Gamma_1\setminus\Delta}),
h_2(a_{\Delta\cap\Gamma_2}\cdot b_{\Gamma_2\setminus\Delta})
\rangle
\in\\
\qquad
\{\langle a_1,a_2\rangle\in\Act_1\times\Act_2:
\forall c_{\Ag\setminus(\Delta\cup\{1,2\})}M,o(w,a_1\cdot a_2\cdot a_{\Delta\setminus(\Gamma_1\cup\Gamma_2)}\cdot c_{\Ag\setminus(\Delta\cup\{1,2\})}\models\chi
\}
\end{array}\]
We denote the subset of $\Act_1\times\Act_2$ above by $A_{1,2,a_{\Delta\setminus(\Gamma_1\cup\Gamma_2)},w,\chi}$. The ability of $\Delta$ to achieve $\chi$ in one step from $w$ is equivalent to the ability of each of $\Delta\cap\Gamma_1$ and $\Delta\cap\Gamma_2$ to enforce actions $a_1$ and $a_2$ on behalf of $1$ and $2$, respectively, so that $\langle a_1,a_2\rangle\in A_{1,2,a_{\Delta\setminus(\Gamma_1\cup\Gamma_2)},w,\chi}$ for some appopriate $a_{\Delta\setminus(\Gamma_1\cup\Gamma_2)}$. Therefore we define $\tr(\atlD{\Delta}\circ\chi)$ as 
\[\bigvee\limits_{a_{\Delta\setminus(\Gamma_1\cup\Gamma_2)}\in\Act_{\Delta\setminus(\Gamma_1\cup\Gamma_2)}}\ \
\bigvee\limits_{A_1\times A_2\subseteq A_{1,2,a_{\Delta\setminus(\Gamma_1\cup\Gamma_2)},w,\chi}}
\atlD{\Delta\cap\Gamma_1}\circ \bigvee\limits_{a_1\in A_1}a_1\wedge
\atlD{\Delta\cap\Gamma_2}\circ \bigvee\limits_{a_2\in A_2}a_2.\]
Formulas obtained by the $\asplit{1}{\Gamma_1}\asplit{2}{\Gamma_2}$-form of $\tr$ are boolean combinations of formulas of the form $\atlD{\Delta}\circ\chi$ where $\Delta\subseteq\Gamma_k$ and $\chi$ is a disjunction of members of $\Act_k$, for $k$ being either $1$ or $2$. In the single $\asplit{.}{.}$ case we are interested in the existence of a satisfying model $\overline{M}$ for $\tr(\varphi)$ as the transitin function $\overline{o}$ of such a model can be used to determine the homomorphism $h$ we need. For the case of $m=2$, the part of $\overline{M}$ is played by a pair of models $\overline{M}_k=\langle\underbrace{\Act_k\cup\{w_{0,k}\}}_{=\overline{W}_k},\langle\overline{\Act}_{k,j}:j\in\Gamma_k\rangle,\overline{o}_k,\overline{V}_k\rangle$ to represent the ability of coalitions withing $\Gamma_k$ to enforce actions with some desired effect on behalf of agent $k$, $k=1,2$. We are interested in the satisfiability of $\tr$-translations at pairs of such models in the following sense. Consider a $\atlD{\Delta}\circ\chi\in\Subf(\tr(\varphi))$ with either $\Delta\subseteq\Gamma_1$ and $\chi$ a boolean combination of atomic propositions from $\overline{AP}_1=\Act_1$, or $\Delta\subseteq\Gamma_2$ and $\chi$ a boolean combination of atomic propositions from $\overline{AP}_2=\Act_2$. We define $\overline{M}_1,\overline{M}_2,w_{0,1},w_{0,2}\models\atlD{\Delta}\circ\chi$ as $\overline{M}_k,w_{0,k}\models\atlD{\Delta}\circ\chi$ for $\psi$ being $\atlD{\Delta}\circ\chi$ with $\Delta\subseteq\Gamma_k$ and $\chi$ written in terms of $\Act_k$, $k=1,2$. The clauses for $\bot$ and for formulas built using $\Rightarrow$ are as usual.

Satisfiability at pair of models of the special type of formulas above straightforwardly reduces to the usual satisfiability at single models once $\tr(\varphi)$ is given a disjunctive normal form: a $\tr(\varphi)$ of this form is satisfiable iff some of its disjunctive members is, and each disjunctive member can be viewed as a conjunction of two formulas $\psi_k$, $\psi_k$ being a conjunction of formulas of the form $\atlD{\Delta}\circ\chi$ with $\Delta\subseteq\Gamma_k$ and $\chi$ written in terms of $\overline{AP}_k$, $k=1,2$. The satisfiability of $\psi_1\wedge\psi_2$ is obviously equivalent to the satisfiability of both $\psi_1$ and $\psi_2$ in the usual sense, at a model of the type of $\overline{M}_k$.

\paragraph{Formulas (\ref{asplitstarflat}) with arbitrary $\asplit{.}{.}$-free $\varphi$} Removing the restriction on $\varphi$s to be in the flat $\atlD{.}\circ$-subset of $\ATL$ makes it necessary to synthesise an $M'$ and the respective $h$ with conditions such as (the many-dimensional form of) (\ref{actprops}) associated with not just one but all the states $w$ of $M$. To enable this, we first elimitate the use of $\until..$ in $\varphi$ using that $|W|$ is known.\footnote{This can cause an $O(|W|)$-blowup in the number of the subformulas of the given $\varphi$, making it clear that we are after nothing more than decidability in principle.} Assuming that $\varphi$ is $\until..$-free, and that $m=1$ again, for the sake of simplicity, we consider assignments $\sem{.}:\Subf(\varphi)\rightarrow 2^W$. We are interested in the existence of an assignment $\sem{.}$ such that an $M'$ that admits a $\Gamma$-to-$i$ homomorphism $h$ to $M$ exists in which $\varphi$ holds at the given state $w$ and $\{w':M',w'\models\psi\}=\sem{\psi}$ for all $\psi\in\Subf(\varphi)$. For $\psi$ being some $p\in AP$ the latter condition holds iff $\sem{p}$ is as detemined from the valuation $V$ of $M$. For $\psi$ being either $\bot$, or with $\Rightarrow$ as the main connective, or of the form $\atlD{\Delta}\circ\psi'$ where $\Delta\cap\Gamma=\emptyset$, $\sem{\psi}$ is similarly unambiguously determined by the identities $\sem{\bot}=\emptyset$, $\sem{\psi'\Rightarrow\psi''}=\sem{\psi'}\Rightarrow\sem{\psi''}$ and $\sem{\atlD{\Delta}\circ\psi'}=\{w'\in W:M,w'\models\atlD{\Delta}\circ\psi'\}$. The latter set can be computed using just $\ATL$ model-checking. Similarly, $\sem{\atlD{\Delta}\circ\psi'}=\{w'\in W:M,w'\models\atlD{(\Delta\setminus\Gamma)\cup\{i\}}\circ\psi'\}$ in case $\Delta\supseteq\Gamma$. Therefore every acceptable assignment is determined unambiguously as soon as its values $\sem{\atlD{\Delta}\circ\psi}$ for $\atlD{\Delta}\circ\psi\in\Subf(\varphi)$ such that $\emptyset\not=\Delta\cap\Gamma\not=\Gamma$ are specified, and the latter values satisfy the inclusions
\[\{w'\in W:M,w'\models\atlD{(\Delta\setminus\Gamma)}\circ\psi'\}\subseteq\sem{\atlD{\Delta}\circ\psi}\subseteq\{w'\in W:M,w'\models\atlD{(\Delta\setminus\Gamma)\cup\{i\}}\circ\psi'\}.\]
Assuming an assignment $\sem{.}$ of the above form, the existence of the required $o'$ and $h$ which link $M'$ to $M$ depends on the satisfiability of the conjunction
\[\bigwedge\limits_{\atlD{\Delta}\circ\psi\in\Subf(\varphi)\atop \emptyset\not=\Delta\cap\Gamma\not=\Gamma}\bigwedge\limits_{w'\in\sem{\atlD{\Delta}\circ\psi}}\bigvee\limits_{a_{\Delta\setminus\Gamma}\in\Act_{\Delta\setminus\Gamma}}\atlD{\Delta\cap\Gamma}\circ\bigvee\limits_{a_i\in A_{i,a_{\Delta\setminus\Gamma},w,\sem{\psi}}} a_i\]
at a model of the type of $\overline{M}$ already introduced above. As expected, here
$A_{i,a_{\Delta\setminus\Gamma},w,\sem{\psi}}=\{a_i\in\Act_i:\forall c_{\Ag\setminus(\Delta\cup\{i\})}(o(w,a_{\Delta\setminus\Gamma}\cdot a_i\cdot c_{\Ag\setminus(\Delta\cup\{i\})})\in X)
\}$.

Obviously the algorithm implied by the above argument is only good to conclude decidability in principle because of the forbidding number of $\sem{.}$s to be considered.

\section{Concluding Remarks}

\paragraph{Related Work}
There is an analogy between our $\asplit{.}{.}$ and the refinement quantifier of {\em Refinement Modal Logic} \cite{DBLP:journals/corr/abs-1202-3538} and its extensions to special classes of multimodal frames \cite{DBLP:conf/aiml/HalesFD12}. \oomit{Refinement $\asplit{i}{\Gamma}$ with $|\Gamma|=2$, which is obviously sufficient to achieve the expressive power of the full language, also bears some analogy with the {\em chop} modality as known from interval logics \cite{Ben,Venema,Zhou}, the related general operator known as {\em bunched implication} \cite{APitts} and the separation modality of Separation Logic \cite{Reynolds}.} Formal studies focusing on controlling the decisions of self-interested delegates can be found in \cite{DBLP:conf/ecai/KrausW12,DBLP:conf/atal/ElkindPW13}.
A notion of {\em refinement} of alternating transition systems, $\ATL$'s original type of models from \cite{AHK97}, allowing, unlike \cite{DBLP:conf/concur/AlurHKV98}, the powers of different {\em sets} of agents to be related, was studied in \cite{DBLP:conf/atal/RyanS01}. The approach of \cite{DBLP:conf/atal/RyanS01} suggests considering a refinement modality of the form $\asplit{\Delta}{\Gamma}$ with $|\Delta|\geq 1$. The authors of \cite{DBLP:conf/atal/RyanS01} stopped short of extending $\ATL$ {\em syntax} by such an operator. Our model-checking algorithm extends to the case of non-singleton coalition-to-coalition refinement as in our CGM-based setting in a straightforward way. Abstraction techniques with the agents being just {\em knowers} were studied in \cite{DBLP:conf/ceemas/EneaD07,DBLP:conf/atal/CohenDLR09}. Abstraction involving over- and under-approximation of coalitions to contain model size was proposed in \cite{DBLP:conf/promas/KosterL11}.  A formalization of teaming sub-agents under a scheduler as turn-based simulation was proposed in \cite{DBLP:conf/atal/GiacomoF10,DBLP:journals/ai/GiacomoPS13}. Modelling varying the considered set of agents is addressed in {\em modular interpreted systems} \cite{DBLP:conf/atal/JamrogaA07,DBLP:journals/corr/JamrogaMS13}. Distinctively, our setting is about varying the set of agents in a system by just redistributing strategic ability, with the overall activities which the system can accommodate unchanged. In CGMs, the effect of actions is defined by means of the transition function. Considering actions which are complete with a description of their effect and an additional parameter to the co-operation modality meant to specify the availability of actions to agents as in \cite{DBLP:conf/lori/HerzigLW13,HerzigPrivate2014} enables specifying delegation too, by varying availability of actions to express their changing hands with their effect on system state being transferred too. This form of delegation is, broadly speaking, complementary to our work as we propose reasoning about migrating the ability to enforce temporal conditions, and {\em synthesizing} implementations in terms of actions through satisfiability checking.
 
\paragraph{Some Work in Progress}
$\asplit{.}{.}$ admits a definition with no reference to $\Gamma$-to-$i$ homomorphisms, which enables translating the $\atlD{.}\circ$-subset of $\ATLA$ into a promising looking subset of many-sorted predicate logic or, similarly, into $\atlD{.}\circ$-subsets of explicit strategy languages such as strategy logics \cite{DBLP:conf/concur/ChatterjeeHP07,DBLP:conf/fsttcs/MogaveroMV10}. Exploring the tractability of the translated formulas is one way of addressing satisfiability in $\ATLA$, which is yet to be done. The translation gives rise to a companion operator, which holds some promise as the means for indirect axiomatization. Regarding direct axiomatization, for any fixed $i$ and $\Gamma$, $\asplit{i}{\Gamma}$ is a ${\bf KD}$- and, with some adjustment to compensate for switching to the local agent vocabulary $\Ag\setminus\{i\}\cup\Gamma$, also a ${\bf T}$-modality. We have also established some non-trivial specific basic equivalences leading to a normal form, and a conventional-looking rule for introducing negative occurrences of $\asplit{.}{.}$, but still lack sufficiently strong axioms for the positive occurrences. 

\paragraph{Acknowledgements}
The research in this paper was partially supported through Bulgarian National Science Fund Grant DID02/32/2009. The author is thankful to the anonymous referees for their careful proof-reading, and to Valentin Goranko, Mark Ryan, Pierre Yves Schobbens and Andreas Herzig for their comments and suggestions.

\bibliographystyle{alpha}
\bibliography{../../bibfiles/mybiblio,del}

\end{document}